\documentclass[10pt,preprint]{aastex}
\usepackage{rotating}

\begin{document}

\title{Microlensing Binaries Discovered through High-Magnification Channel}

\author{
I.-G. Shin\altaffilmark{1},       
J.-Y. Choi\altaffilmark{1},       
S.-Y. Park\altaffilmark{1},       
C. Han\altaffilmark{1,72,79},
A. Gould\altaffilmark{9,72},    
T. Sumi\altaffilmark{30,73},        
A. Udalski\altaffilmark{31,74},    
J.-P. Beaulieu\altaffilmark{34,75}, 
M. Dominik\altaffilmark{47,76,77,78},\\
and\\
W. Allen\altaffilmark{2},
M. Bos\altaffilmark{3},
G.W. Christie\altaffilmark{4},   
D.L. Depoy\altaffilmark{5},      
S. Dong\altaffilmark{6}, 
J. Drummond\altaffilmark{7},
A. Gal-Yam\altaffilmark{8},
B.S. Gaudi\altaffilmark{9},      
L.-W. Hung\altaffilmark{10},
J. Janczak\altaffilmark{11},
S. Kaspi\altaffilmark{12},
C.-U. Lee\altaffilmark{13},      
F. Mallia\altaffilmark{14}
D. Maoz\altaffilmark{12},
A. Maury\altaffilmark{14},
J. McCormick\altaffilmark{15},    
L.A.G. Monard\altaffilmark{16},
D. Moorhouse\altaffilmark{17},    
J. A. Mu\~{n}oz\altaffilmark{18},
T. Natusch\altaffilmark{4},
C. Nelson\altaffilmark{19},
B.-G. Park\altaffilmark{13}, 
R.W.\ Pogge\altaffilmark{9},      
D. Polishook\altaffilmark{12},    
Y. Shvartzvald\altaffilmark{12},
A. Shporer\altaffilmark{12},    
G. Thornley\altaffilmark{17},
J.C. Yee\altaffilmark{9}\\
(The $\mu$FUN Collaboration),\\
F. Abe\altaffilmark{20},        
D.P. Bennett\altaffilmark{21},  
I.A. Bond\altaffilmark{22},
C.S. Botzler\altaffilmark{23},   
A. Fukui\altaffilmark{20},   
K. Furusawa\altaffilmark{20},   
F. Hayashi\altaffilmark{20},    
J.B. Hearnshaw\altaffilmark{24}, 
S. Hosaka\altaffilmark{20},     
Y. Itow\altaffilmark{20},        
K. Kamiya\altaffilmark{20},      
P.M. Kilmartin\altaffilmark{25}, 
S. Kobara\altaffilmark{20},
A. Korpela\altaffilmark{26},     
W. Lin\altaffilmark{22},         
C.H. Ling\altaffilmark{22},     
S. Makita\altaffilmark{20},      
K. Masuda\altaffilmark{20},      
Y. Matsubara\altaffilmark{20},    
N. Miyake\altaffilmark{20},
Y. Muraki\altaffilmark{27},     
M. Nagaya\altaffilmark{20},
K. Nishimoto\altaffilmark{20},   
K. Ohnishi\altaffilmark{28},    
T. Okumura\altaffilmark{20},
K. Omori\altaffilmark{20},
Y.C. Perrott\altaffilmark{23},   
N. Rattenbury\altaffilmark{23},   
To. Saito\altaffilmark{29},    
L. Skuljan\altaffilmark{22},     
D.J. Sullivan\altaffilmark{26},  
D. Suzuki\altaffilmark{30},      
W.L. Sweatman\altaffilmark{22}, 
P.J. Tristram\altaffilmark{25},  
K. Wada\altaffilmark{30},        
P.C.M. Yock\altaffilmark{23}\\   
(The MOA Collaboration),\\
M.K. Szyma\'nski\altaffilmark{31},  
M. Kubiak\altaffilmark{31},            
G. Pietrzy\'nski\altaffilmark{31,32}, 
I. Soszy\'nski\altaffilmark{31},      
R. Poleski\altaffilmark{31},       
K.\ Ulaczyk\altaffilmark{31},
{\L}. Wyrzykowski\altaffilmark{31,33},
S. Koz{\l}owski\altaffilmark{31},
P. Pietrukowicz\altaffilmark{31}\\
(The OGLE Collaboration)\\
M.D. Albrow\altaffilmark{24},
V. Batista\altaffilmark{9},    
D.M. Bramich\altaffilmark{46},
S. Brillant\altaffilmark{35},     
J.A.R. Caldwell\altaffilmark{69},
J.J. Calitz\altaffilmark{71},
A. Cassan\altaffilmark{34},       
A. Cole\altaffilmark{36},         
K.H. Cook\altaffilmark{70},
E. Corrales\altaffilmark{34},      
Ch. Coutures\altaffilmark{34},     
S. Dieters\altaffilmark{34,37},    
D. Dominis Prester\altaffilmark{38},
J. Donatowicz\altaffilmark{39},
P. Fouqu\'e\altaffilmark{37},     
J. Greenhill\altaffilmark{36},     
M. Hoffman\altaffilmark{71},
U.G. J{\o}rgensen\altaffilmark{60,61},
S. R. Kane\altaffilmark{40},
D. Kubas\altaffilmark{34,35},
J.-B. Marquette\altaffilmark{34},
R. Martin\altaffilmark{44},
P. Meintjes\altaffilmark{71},
J. Menzies\altaffilmark{41},
K.R. Pollard\altaffilmark{24}, 
K. C. Sahu\altaffilmark{42},
J. Wambsganss\altaffilmark{43},
A. Williams\altaffilmark{44},
C. Vinter\altaffilmark{60},
M. Zub\altaffilmark{43}\\
(The PLANET Collaboration)\\
A. Allan\altaffilmark{45},
P. Browne\altaffilmark{47}, 
K. Horne\altaffilmark{47},          
C. Snodgrass\altaffilmark{48,35},  
I. Steele\altaffilmark{49},  
R. Street\altaffilmark{50},       
Y. Tsapras\altaffilmark{50}\\
(The RoboNet Collaboration)\\
and\\
K.A. Alsubai\altaffilmark{51},
V. Bozza\altaffilmark{52},
P. Browne\altaffilmark{47},
M.J. Burgdorf\altaffilmark{53,54},
S. Calchi Novati\altaffilmark{52,55},
P. Dodds\altaffilmark{47},
S. Dreizler\altaffilmark{56},
F. Finet\altaffilmark{57},
T. Gerner\altaffilmark{58},
M. Glitrup\altaffilmark{59},
F. Grundahl\altaffilmark{59},
S. Hardis\altaffilmark{60},
K. Harps{\o}e\altaffilmark{60,61},
F.V. Hessman\altaffilmark{56},
T.C. Hinse\altaffilmark{13,60,62},
M. Hundertmark\altaffilmark{47,56},
N. Kains\altaffilmark{47,63},
E. Kerins\altaffilmark{64},
C. Liebig\altaffilmark{47,58},
G. Maier\altaffilmark{58},
L. Mancini\altaffilmark{52,65},
M. Mathiasen\altaffilmark{60},
M.T. Penny\altaffilmark{64},
S. Proft\altaffilmark{58},
S. Rahvar\altaffilmark{66},
D. Ricci\altaffilmark{57},
G. Scarpetta\altaffilmark{52,67},
S. Sch\"{a}fer\altaffilmark{56},
F. Sch\"{o}nebeck\altaffilmark{58},
J. Skottfelt\altaffilmark{60},
J. Surdej\altaffilmark{57},
J. Southworth\altaffilmark{68},
F. Zimmer\altaffilmark{58}\\ 
(The MiNDSTEp Consortium)\\
}

\altaffiltext{1}{Department of Physics, Institute for Astrophysics, Chungbuk National University, Cheongju 371-763, Korea}
\altaffiltext{2}{Vintage Lane Observatory, Blenheim, New Zealand}
\altaffiltext{3}{Molehill Astronomical Observatory, North Shore, New Zealand}
\altaffiltext{4}{Auckland Observatory, P.O. Box 24-180, Auckland, New Zealand} 
\altaffiltext{5}{Department of Physics, Texas A\&M University, College Station, TX, USA} 
\altaffiltext{6}{Institute for Advanced Study, Einstein Drive, Princeton, NJ 08540, USA}
\altaffiltext{7}{Possum Observatory, Patutahi, New Zealand}
\altaffiltext{8}{Benoziyo Center for Astrophysics, the Weizmann Institute, Israel} 
\altaffiltext{9}{Department of Astronomy, Ohio State University, 140 W. 18th Ave., Columbus, OH 43210, USA}
\altaffiltext{10}{Department of Physics \& Astronomy, University of California Los Angeles, Los Angeles, CA 90095, USA}
\altaffiltext{11}{Department of Physics, Ohio State University, 191 W. Woodruff, Columbus, OH 43210, USA}
\altaffiltext{12}{School of Physics and Astronomy, Tel-Aviv University, Tel Aviv 69978, Israel} 
\altaffiltext{13}{Korea Astronomy and Space Science Institute, Daejeon 305-348, Korea} 
\altaffiltext{14}{Campo Catino Austral Observatory, San Pedro de Atacama, Chile}
\altaffiltext{15}{Farm Cove Observatory, Pakuranga, Auckland} 
\altaffiltext{16}{Bronberg Observatory, Pretoria, South Africa} 
\altaffiltext{17}{Kumeu Observatory, Kumeu, New Zealand} 
\altaffiltext{18}{Departamento de Astronomi{\'a} y Astrof{\'i}sica, Universidad de Valencia, E-46100 Burjassot, Valencia, Spain}
\altaffiltext{19}{College of Optical Sciences, University of Arizona, 1630 E. University Blvd, Tucson Arizona, 85721, USA}
\altaffiltext{20}{Solar-Terrestrial Environment Laboratory, Nagoya University, Nagoya, 464-8601, Japan}
\altaffiltext{21}{Department of Physics, University of Notre Damey, Notre Dame, IN 46556, USA}
\altaffiltext{22}{Institute of Information and Mathematical Sciences, Massey University, Private Bag 102-904, North Shore Mail Centre, Auckland, New Zealand} 
\altaffiltext{23}{Department of Physics, University of Auckland, Private Bag 92019, Auckland, New Zealand}
\altaffiltext{24}{University of Canterbury, Department of Physics and Astronomy, Private Bag 4800, Christchurch 8020, New Zealand}  
\altaffiltext{25}{Mt. John Observatory, P.O. Box 56, Lake Tekapo 8770, New Zealand} 
\altaffiltext{26}{School of Chemical and Physical Sciences, Victoria University, Wellington, New Zealand} 
\altaffiltext{27}{Department of Physics, Konan University, Nishiokamoto 8-9-1, Kobe 658-8501, Japan} 
\altaffiltext{28}{Nagano National College of Technology, Nagano 381-8550, Japan} 
\altaffiltext{29}{Tokyo Metropolitan College of Industrial Technology, Tokyo 116-8523, Japan} 
\altaffiltext{30}{Department of Earth and Space Science, Osaka University, Osaka 560-0043, Japan}
\altaffiltext{31}{Warsaw University Observatory, Al. Ujazdowskie 4, 00-478 Warszawa, Poland}
\altaffiltext{32}{Universidad de Concepci\'on, Departamento de Fisica, Casilla 160-C, Concepci{\'o}n, Chile} 
\altaffiltext{33}{Institute of Astronomy Cambridge University, Madingley Road, CB3 0HA Cambridge, UK} 
\altaffiltext{34}{Institut d'Astrophysique de Paris, UMR7095 CNRS--Universit{\'e} Pierre \& Marie Curie, 98 bis boulevard Arago, 75014 Paris, France} 
\altaffiltext{35}{European Southern Observatory, Casilla 19001, Vitacura 19, Santiago, Chile} 
\altaffiltext{36}{School of Math and Physics, University of Tasmania, Private Bag 37, GPO Hobart, Tasmania 7001, Australia} 
\altaffiltext{37}{LATT, Universit\'e de Toulouse, CNRS, 14 Avenue Edouard Belin, 31400 Toulouse, France} 
\altaffiltext{38}{Physics Department, Faculty of Arts and Sciences, University of Rijeka, Omladinska 14, 51000 Rijeka, Croatia}
\altaffiltext{39}{Technical University of Vienna, Department of Computing, Wiedner Hauptstrasse 10, Vienna, Austria}
\altaffiltext{40}{NASA Exoplanet Science Institute, Caltech, MS 100-22, 770 South Wilson Avenue, Pasadena, CA 91125, USA}
\altaffiltext{41}{South African Astronomical Observatory, P.O. Box 9 Observatory 7935, South Africa} 
\altaffiltext{42}{Space Telescope Science Institute, 3700 San Martin Drive, Baltimore, MD 21218, USA}
\altaffiltext{43}{Astronomisches Rechen-Institut (ARI), Zentrum f{\"u}r Astronomie der Universit{\"a}t Heidelberg (ZAH), M{\"o}nchhofstrasse 12-14, 69120 Heidelberg, Germany}
\altaffiltext{44}{Perth Observatory, Walnut Road, Bickley, Perth 6076, Australia}
\altaffiltext{45}{School of Physics, University of Exeter, Stocker Road, Exeter, Devon, EX4 4QL, UK}
\altaffiltext{46}{European Southern Observatory, Karl-Schwarzschild-Stra{\ss}e 2, 85748 Garching bei M{\"u}nchen, Germany}
\altaffiltext{47}{School of Physics \& Astronomy, SUPA, University of St. Andrews, North Haugh, St. Andrews, KY16 9SS, UK}
\altaffiltext{48}{Max-Planck-Institut f{\"o}r Sonnensystemforschung, Max-Planck-Str. 2, 37191 Katlenburg-Lindau, Germany}
\altaffiltext{49}{Astrophysics Research Institute, Liverpool John Moores University, Egerton Wharf, Birkenhead CH41 1LD, UK}
\altaffiltext{50}{Las Cumbres Observatory Global Telescope Network, 6740B Cortona Dr, Suite 102, Goleta, CA 93117, USA} 
\altaffiltext{51}{Qatar Foundation, P.O. Box 5825, Doha, Qatar}
\altaffiltext{52}{Universit\`{a} degli Studi di Salerno, Dipartimento di Fisica ``E.R. Caianiello'', Via S. Allende, 84081 Baronissi (SA), Italy}
\altaffiltext{53}{Deutsches SOFIA Institut, Universit\"{a}t Stuttgart, Pfaffenwaldring 31, 70569 Stuttgart, Germany}
\altaffiltext{54}{SOFIA Science Center, NASA Ames Research Center, Mail Stop N211-3, Moffett Field CA 94035, USA}
\altaffiltext{55}{Istituto Internazionale per gli Alti Studi Scientifici (IIASS), Vietri Sul Mare (SA), Italy}
\altaffiltext{56}{Institut f\"{u}r Astrophysik, Georg-August-Universit\"{a}t, Friedrich-Hund-Platz 1, 37077 G\"{o}ttingen, Germany}
\altaffiltext{57}{Institut d'Astrophysique et de G\'{e}ophysique, All\'{e}e du 6 Ao\^{u}t 17, Sart Tilman, B\^{a}t.\ B5c, 4000 Li\`{e}ge, Belgium}
\altaffiltext{58}{Astronomisches Rechen-Institut, Zentrum f\"{u}r Astronomie der Universit\"{a}t Heidelberg (ZAH),  M\"{o}nchhofstr.\ 12-14, 69120 Heidelberg, Germany}
\altaffiltext{59}{Department of Physics \& Astronomy, Aarhus Universitet, Ny Munkegade, 8000 {\AA}rhus C, Denmark}
\altaffiltext{60}{Niels Bohr Institute, University of Copenhagen, Juliane Maries vej 30, 2100 Copenhagen, Denmark}
\altaffiltext{61}{Centre for Star and Planet Formation, Geological Museum, {\O}ster Voldgade 5, 1350 Copenhagen, Denmark}
\altaffiltext{62}{Armagh Observatory, College Hill, Armagh, BT61 9DG, Northern Ireland, UK}
\altaffiltext{63}{ESO Headquarters, Karl-Schwarzschild-Str. 2, 85748 Garching bei M\"{u}nchen, Germany}
\altaffiltext{64}{Jodrell Bank Centre for Astrophysics, University of Manchester, Oxford Road,Manchester, M13 9PL, UK}
\altaffiltext{65}{Max Planck Institute for Astronomy, K\"{o}nigstuhl 17, 619117 Heidelberg, Germany}
\altaffiltext{66}{Department of Physics, Sharif University of Technology, P.~O.\ Box 11155--9161, Tehran, Iran}
\altaffiltext{67}{INFN, Gruppo Collegato di Salerno, Sezione di Napoli, Italy}
\altaffiltext{68}{Astrophysics Group, Keele University, Staffordshire, ST5 5BG, UK}
\altaffiltext{69}{McDonald Observatory, 16120 St Hwy Spur 78 \#2, Fort Davis, TX 79734, USA}
\altaffiltext{70}{Institute of Geophysics and Planetary Physics (IGPP), L-413, Lawrence Livermore National Laboratory, PO Box 808, Livermore, CA 94551, USA}
\altaffiltext{71}{University of the Free State, Faculty of Natural and Agricultural Sciences, Department of Physics, PO Box 339, Bloemfontein 9300, South Africa}
\altaffiltext{72}{The $\mu$FUN Collaboration}
\altaffiltext{73}{The MOA Collaboration}
\altaffiltext{74}{The OGLE Collaboration}
\altaffiltext{75}{The PLANET Collaboration}
\altaffiltext{76}{The RoboNet Collaboration}
\altaffiltext{77}{The MiNDSTEp Consortium}
\altaffiltext{78}{Royal Society University Research Fellow}
\altaffiltext{79}{Corresponding author}

\begin{abstract}
Microlensing can provide a useful tool to probe binary distributions 
down to low-mass limits of binary companions.  In this paper, we 
analyze the light curves of 8 binary lensing events detected through 
the channel of high-magnification events during the seasons from 2007 
to 2010.  The perturbations, which are confined near the peak of the 
light curves, can be easily distinguished from the central perturbations 
caused by planets.  However, the degeneracy between close and wide 
binary solutions cannot be resolved with a $3\sigma$ confidence level 
for 3 events, implying that the degeneracy would be an important 
obstacle in studying binary distributions.  The dependence of the 
degeneracy on the lensing parameters is consistent with a theoretic 
prediction that the degeneracy becomes severe as the binary separation 
and the mass ratio deviate from the values of resonant caustics.  
The measured mass ratio of the event OGLE-2008-BLG-510/MOA-2008-BLG-369 
is $q\sim 0.1$, making the companion of the lens a strong brown-dwarf 
candidate.
\end{abstract}

\keywords{gravitational lensing: micro -- binaries: general}

\section{Introduction}

Microlensing can be used to probe the distributions of binary companions 
of Galactic stars as functions of mass ratio and separation, which provide 
important observational constraints on theories of star formation. Being 
sensitive to low-mass companions that are difficult to be detected by 
other methods, microlensing enables to make complete distributions down 
to the low mass limit of binary companions \citep{gould01}.

Despite the importance, the progress of this application of microlensing 
to the statistical analysis of binaries has been stagnant. There are two 
main reasons for this. The first reason arises due to the difficulties 
in estimating the detection efficiency of binary lenses. Previously, 
lensing events caused by binary lenses were mainly detected through 
accidental detections of sudden rises and falls of the source flux 
resulting from source crossings over caustics formed by binary lenses, 
e.g.\ \citet{udalski94}, \citet{alcock00}, \citet{jaroszynski04, 
jaroszynski06, jaroszynski10}, and \citet{skowron07}.  The caustics 
represent the positions on the source plane at which the lensing 
magnification of a point source becomes infinite. For binary events 
detected through this channel, it is difficult to estimate the detection 
efficiency due to the haphazard nature of caustic crossings.  The second 
reason is that microlensing is mainly sensitive to binaries distributed 
over a narrow range of separations. The probability of caustic crossings 
increases with the increase of the caustic size. The caustic size becomes 
maximum when the separation between the lens components is of order the 
Einstein radius, $\theta_{\rm E}$, and decreases rapidly with the increase 
or decrease of the separation from $\theta_{\rm E}$. As a result, the 
majority of microlensing binaries have separations distributed within a 
small range. This limits especially the study of the distribution of binary 
separations.

However, under the current observational strategy of microlensing experiments 
focusing on planet detections, a significant fraction of binary events are 
detected through a new channel of high-magnification events. For the detections 
of short-duration planetary signals in lensing light curves, planetary lensing 
experiments are being conducted in survey and follow-up mode, where alerts of 
ongoing events are issued by survey experiments and intensive observations of 
these events are conducted by follow-up experiments. In this mode, high-magnification 
events are the most important targets for follow-up observations because the source 
trajectories of these events always pass close to the central perturbation region 
induced by the planet and thus the efficiency of planet detections is very high 
\citep{griest98}.  In addition, the time of the perturbation can be predicted in 
advance and thus intensive follow-up can be prepared.  This leads to an observational 
strategy of intensively monitoring all high-magnification events regardless of 
whether they show signals of planets.

In addition to planets, high-magnification events are sensitive to binaries as 
well, especially those with separations substantially smaller (close binaries) 
or larger (wide binaries) than the Einstein radius. For close binaries, there 
exist three caustics where one is formed around the center of mass of the binary 
and the other two are located away from the barycenter. For wide binaries, on 
the other hand, there exist two caustics each of which is located adjacent to the 
individual lens components. Then, high-magnification events resulting from the 
source trajectories passing either close to the center of mass of a close binary 
or one of the components of a wide binary are sensitive to binaries. The high 
sensitivity to close and wide binaries combined with the strategy of monitoring 
all high-magnification events imply that binary events detected through the 
high-magnification channel are important for the construction of an unbiased 
sample of binaries with a wider range of separations and thus for the statistical 
studies of binaries \citep{han09b}.

In this paper, we analyze the light curves of 8 binary microlensing events 
detected through the high-magnification channel during the seasons from 2007 
to 2010. We search for the solutions of binary lensing parameters by conducting 
modeling of the light curves.  We discuss the characteristics of the binaries.

\section{Observation}

All 8 tested events analyzed in this work were detected toward the Galactic 
bulge direction. In Table 1, we list the coordinates of the events.  Each 
event is designated first by the microlensing group who first discovered 
the event and then followed by the year when the event was discovered. If 
an event is discovered independently by two different groups, they are named 
separately. For example, the event OGLE-2008-BLG-510/MOA-2008-BLG-368 was 
discovered by both OGLE and MOA groups in 2008.  For all events, the peak 
magnifications are high and thus they are issued as important targets for 
follow-up observations by the MOA \citep{bond01,sumi03} and OGLE \citep{udalski03} 
survey experiments.  As a result, the peaks of the light curves were densely 
covered by follow-up observations including the $\mu$FUN \citep{gould06}, 
PLANET \citep{beaulieu06}, RoboNet \citep{tsapras09}, and MiNDSTEp 
\citep{dominik10}.  In Table 2, we list the survey and follow-up groups who 
participated in the observation of the individual events. In Table 3, we 
also list the telescopes used for observations along with their locations.

The photometry of the data was conducted by using the codes developed by 
the individual groups.  For some events, we re-reduced data based on the 
image subtraction method to ensure better photometry.  The error bars of 
the data sets were rescaled so that $\chi^2/{\rm dof}$ becomes unity for 
the data set of each observatory where $\chi^2$ is computed based on the 
best-fit model.

In Figure 1 -- 8, we present the light curves of the individual events. 
For all events, the common feature of the light curves is that most of 
the light curve is consistent with the standard single-lens light curve 
\citep{paczynski86} and the perturbation is confined in a narrow region 
around the peak.

\section{Modeling}

For the light curve of each event, we search for solutions of lensing parameters 
in the space encompassing both stellar and planetary companions.  The light 
curve of a binary-lens event is characterized by 6 basic parameters.  The 
first 3 parameters are related to the geometry of the lens-source approach.  
They are the Einstein time scale, $t_{\rm E}$, the time of the closest lens-source 
approach, $t_0$, and the lens source separation at that moment, $u_0$. The other 
3 parameters are related to the binarity of the lens. These parameters are the 
mass ratio between the lens components, $q$, the projected separation in units 
of the Einstein radius, $s$, and the angle between the source trajectory and 
the binary axis, $\alpha$. For all tested events, the perturbations exhibit 
features caused either by crossings over or approaches close to caustics and 
thus it is required to consider the modification of magnifications caused by 
the finite-source effect during the perturbation. This requires to include an 
additional parameter of the normalized source radius, $\rho_\star$, which is 
related to the angular source radius, $\theta_\star$, and the Einstein radius 
by $\rho_\star=\theta_\star/\theta_{\rm E}$.

For each event, we search for the solution of the best-fit parameters by minimizing 
$\chi^2$ in the parameter space. We do this by dividing the parameters into two 
categories. For the parameters in the first category, grid searches are conducted. 
For the remaining parameters in the second category are searched by using a downhill 
approach. We choose $s$, $q$, and $\alpha$ as the grid parameters because these 
parameters are related to the features of lensing light curves in a complicated 
pattern while the other parameters are more directly related to the features of 
the light curve. For the $\chi^2$ minimization, we use a Markov Chain Monte Carlo 
method. Brute-force search over the space of the grid parameters is needed in order 
to investigate possible local minima of degenerate solutions. This is important 
because it is known that there exists a pair of close/wide solution for binary-lens 
events, especially for binaries with separations substantially smaller or larger 
than the Einstein radius \citep{dominik99}.  Once local minima are identified, 
we check all of them by gradually narrowing down the grid parameter space.  When 
the space is sufficiently confined, we allow the grid parameters to vary in order 
to pin down the exact location of the solution.

Computation of magnifications affected by the finite-source effect is based on 
the ray-shooting method \citep{schneider86, kayser86, wambsganss97}.  In this 
numerical method, rays are uniformly shot from the image plane, bent according 
to the lens equation, and land on the source plane. Then, the finite magnification 
is computed by comparing the number densities of rays on the image and source 
planes.  This method requires heavy computation because a large number of rays are 
needed for accurate magnification computation. We accelerate the computation 
by using two major methods.  The first method is applying the ``map making'' method 
\citep{dong06}. In this method, a map for a given set of $(s,q)$ is used to produce 
numerous light curves resulting from different source trajectories instead of 
shooting rays all over again. The second method is applying the semi-analytic 
hexadecapole approximation \citep{pejcha09, gould08} for the finite magnification 
computation when the source is not very close to the caustic.

In computing finite magnifications, we consider the effect of limb-darkening of 
the source star surface by modeling the surface brightness by
\begin{equation}
S_{\lambda} = {{F_{\lambda}}\over{\pi \theta^2 _{\star}}} 
{ { \left[ 1- \Gamma_{\lambda} \left( 1- {3\over2} \cos\phi \right) \right]} }
\end{equation}
where $\Gamma_{\lambda}$ is the linear limb-darkening coefficient,  $F_{\lambda}$ 
is the flux from the source star, and $\phi$ is the angle between the line of 
sight toward the source star and the normal to the source star's surface. We 
choose the coefficients from \citet{claret00}, where the source type is determined 
from the location of the source star on the color-magnitude diagram. In Table 
\ref{table:four}, we present the coefficients of the individual events.

In addition to the modeling based on standard binary-lensing parameters, we 
conduct modeling considering the second-order effects on the light curve.
The first effect is the ``parallax effect'' that is caused by the change of 
the observer position induced by the orbital motion of the Earth around the 
Sun \citep{gould92, alcock95}.  The second effect is the ``orbital effect'' 
caused by the change of the lens position induced by the orbital motion of the 
lens \citep{albrow02, shin11, skowron11}.  Measurement of the parallax effect 
is important because it allows to determine the physical parameters of the lens 
system \citep{gould92}.  Detecting the orbital effect is important because it 
can help to characterize the orbital parameters of the lens system.

\section{Results}

In Table 5, we present the best-fit parameters found from modeling.  For 
each event, we present the pair of close and wide binary solutions in order 
to show the severity of the degeneracy.  The best-fit light curves of the 
individual events are overplotted on the data in Figure 1 -- 8.  In Figure 
\ref{fig:nine}, we also present the geometry of the lens systems.  For each 
event, we present two sets of geometry corresponding to the close (left panel) 
and wide (right panel) binary solutions.  In each panel, the big and small 
dots represent the locations of the binary lens components with heavier and 
lighter masses, respectively.  The closed figure with cusps represents the 
caustic and the straight line with an arrow represents the source trajectory 
with respect to the caustic. The empty circle near the tip of the arrow on 
the source trajectory represents the source size. The dashed circle represents 
the Einstein ring.  For the close binary, there exists a single Einstein ring 
whose radius corresponds to the total mass of the binary.  For the wide binary, 
on the other hand, there exist two rings with radii corresponding to the masses 
of the individual lens components.  The small panel on the right side of each 
main panel shows the enlargement of the region around the caustic.  We find 
that the perturbations of the events MOA-2008-BLG-159, MOA-2009-BLG-408, 
MOA-2010-BLG-349, and MOA-2010-BLG-546 were produced by the source star's 
crossing over the central caustic.  For the events MOA-2007-BLG-146, 
OGLE-2008-BLG-510/MOA-2008-BLG369, MOA-2010-BLG-266, and MOA-2010-BLG-406, 
on the other hand, the perturbations were produced by the approach of the 
source trajectory close to one of the cusps of the central caustic.

We find that the modeling including the parallax and orbital effects does 
not yield solutions with statistically significant $\chi^2$ improvement.  
Considering that the range of the time scales of the events is $5\ {\rm days} 
\lesssim t_{\rm E}\lesssim 30\ {\rm days}$, we judge that the difficulties 
in detecting the second-order effects are due to the short time scales of 
the events.  Since the lens parallaxes are not measured, we are not able to 
determine the physical parameters of lenses. However, for 5 events we are 
able to measure the Einstein radii, which is another quantity to constrain 
the physical lens parameters.  The Einstein radius is measured from the 
deviation of the light curve caused by the finite-source effect.  By detecting 
the deviation, the normalized source radius $\rho_\star$ is measured from 
modeling.  With the additional information of the source radius, which is 
obtained from the location of the source star on the color-magnitude diagram 
of stars in the field around the source star, the Einstein radius is determined 
as $\theta_{\rm E}= \theta_\star/\rho_\star$ \citep{yoo04}. With the measured 
Einstein radius, the relative lens-source proper motion is determined by 
$\mu = \theta_{\rm E}/ t_{\rm E}$. The values of the measured Einstein radii 
and the proper motions are presented in Table \ref{table:five}.  Among the 5 
events for which the Einstein radius is measured, 4 events are caustic-crossing 
events.  For the case of MOA-2007-BLG-146, the center of the source star did 
not cross the caustic but the edge of the source passed over the caustic and 
thus the Einstein radius was measurable.

It is known that central perturbations, which are the common features for all 
analyzed events, can be produced either by planetary companions or binaries 
\citep{albrow02, han08, han09a, hankim09}.  We find that the planet/binary 
degeneracy is easily distinguished and the binary origin can be firmly identified. 
The range of the mass ratios is $0.1\lesssim q\lesssim 0.73$.\footnote{In Table 
\ref{table:five}, the value of the mass ratio $q>1$ represents the case where 
the source trajectory approaches the lighter component of the binary.} We note 
that the event OGLE-2008-BLG-510/MOA-2008-BLG-369 is caused by a binary with 
a low-mass companion.  Although the absolute value of the lens mass cannot 
be determined, the measured mass ratio $q\sim 0.1$ makes the companion of 
the binary a brown-dwarf candidate considering that the time scale of the 
event $t_{\rm E}\sim 27$ days is a typical one for Galactic bulge events 
caused by low-mass stars.  Therefore, this event demonstrates that microlensing 
is a useful tool to study low-mass binary companions including brown dwarfs.  
By the time of completing this paper, we learned that \citet{bozza11} 
released the result of analysis for OGLE-2008-BLG-510/MOA-2008-BLG-369.  
Their result is very consistent with ours and stated the possibility of 
the brown dwarf companion.

Although the binary nature of the lenses is clearly identified, it is found 
that the degeneracy between the close and wide binary solutions is severe 
for some events. The close/wide binary degeneracy, which results from a 
symmetry in the lens equation, was first mentioned by \citet{griest98} and 
further investigated by \citet{dominik99}. The events for which the degeneracy 
cannot be distinguished with a $3\sigma$ confidence level include 
OGLE-2008-BLG-510/MOA-2008-BLG-369, MOA-2009-BLG-408, and MOA-2010-BLG-546.  
The severity of the degeneracy and the correspondence in the lens-system 
geometry between the pairs of degenerate solutions can be seen from the 
comparison of the geometry of the lens system at the time of perturbation.  
As predicted by theoretical studies, the close/wide degeneracy is caused by 
the similarity of the shape between the caustics of the close and wide binaries  
The caustic shape results from the combination of the projected separation and 
mass ratio.  To see how the severity of the degeneracy depends on these parameters, 
we plot the locations of the degenerate solutions in the parameter space of $s$ 
and $q$ in Figure \ref{fig:ten}. In the plot, the filled dots denote that the 
degeneracy is resolved at the $3\sigma$ confidence level and the empty dots 
symbolize that the degeneracy is not resolved. The area encompassed by dashed 
lines represents the region within which the lens forms a single merged large 
caustic (resonant caustic).  From the plot, it is found that the degeneracy 
becomes severe as the binary separation is located well away from the range 
of resonant caustics.  Therefore,  the degeneracy would be an important 
obstacle in studying binary distributions for binaries with very close or 
wide separations.

\section{Conclusion and Discussion}

We conducted modeling of light curves of 8 binary lensing events detected 
through the high-magnification channel during 2007 -- 2010 seasons. We 
found that the binary/planet degeneracy of the central perturbations were 
easily distinguished. However, the degeneracy between the close and wide 
binary solutions could not be resolved with confidence for some of the 
events. We confirmed the theoretic prediction that the degeneracy becomes 
severe for binaries with separations substantially smaller or wider than 
the Einstein radius and thus the close/wide degeneracy would be an important 
obstacle in the studies of binary distributions. For one of the events, 
the measured mass ratio is in the range of a brown dwarf, demonstrating 
that microlensing is a useful tool to study low-mass binary companions.

Although it is difficult to draw meaningful statistical properties of 
binaries based on the handful events analyzed in this work, it is expected 
that the microlensing use of binary statistics would expand.  One way for 
this improvement is the removal of human intervention in the selection process 
of a follow-up campaign. An example of this effort is the SIGNALMEN anomaly 
detector achieved by the ARTEMiS system \citep{dominik07}.  Another way is 
conducting high-cadence surveys to dispense with follow-up observations. 
Recently, the OGLE group significantly increased the observational cadence 
by upgrading its camera with a wider field of view to the level of being 
able to detect short planetary perturbations by the survey itself.  The 
Korea Microlensing Telescope Network (KMTNet) is a planned survey experiment 
that will achieve ~10 minute sampling of all lensing events by using a network 
of 1.6 m telescopes to be located in three different continents in the Southern
hemisphere with wide-field cameras.  These new type surveys will enable not 
only to densely cover events but also to significantly increase the number 
of events in binary samples.  Being able to detect and densely cover binary 
events without human intervention combined with the increased number of events
will enable microlensing to become a useful method to study binary statistics.

Even with the increase of the number of events and the improvement of the 
process of obtaining samples, it is still an important issue to resolve the 
close/wide degeneracy. \citet{han99} proposed that astrometric observation 
of the centroid motion of a lensed star by using a high-resolution instrument 
makes it possible to resolve the ambiguity of the photometric binary-lens fit 
for most accidentally degenerate cases.  However, it is found that the close/wide 
binary degeneracy is so severe that it causes the image centroids of the wide and 
close solutions to follow a similar pattern of motion although the motions of the 
image centroid for the two degenerate cases are displaced from one another long 
after the event and thus the degeneracy can eventually be resolvable \citep{han00}. 
In addition, this method requires space-based astrometric instrument and thus 
can not be applicable to events being detected by current lensing experiments. 
A class of events for which the degeneracy can be photometrically resolved
are repeating events where the source trajectory passes both the central 
perturbation region of one of the binary components and the effective lensing 
region of the other binary component, e.g. OGLE-2009-BLG-092/MOA-2009-BLG-137 
\citep{ryu10}. However, this method can be applicable to a small fraction of 
events.
%
Therefore, devising a general method resolving this degeneracy would be crucial 
for the statistical binary studies of microlensing binaries.

\acknowledgments 
Work by CH was supported by Creative Research Initiative 
Program (2009-0081561) of National Research Foundation of Korea.
The OGLE project has received funding from the European Research
Council under the European Community's Seventh Framework Programme
(FP7/2007-2013) / ERC grant agreement no. 246678.
Work by BSG and AG was supported in part by NSF grant AST-1103471.
Work by BSG, AG, RWP, and JCY supported in part by NASA grant NNX08AF40G.
Work by JCY was supported by a National Science Foundation Graduate
Research Fellowship under Grant No.\ 2009068160.
Work by MH was supported by Qatar National Research Fund and Deutsche Forschungsgemeinschaft.
The MOA experiment was supported by JSPS22403003, JSPS20340052, JSPS18253002, and JSPS17340074.
TS was supported by the grants JSPS18749004, MEXT19015005, and JSPS20740104.
FF, DR and JS acknowledge was supported by the Communaut{\'e}
fran\c{c}aise de Belgique - Actions de recherche concert{\'e}es -
Acad{\'e}mie universitaire Wallonie-Europe.

\clearpage

\begin{figure}[ht]
\epsscale{0.8}
\plotone{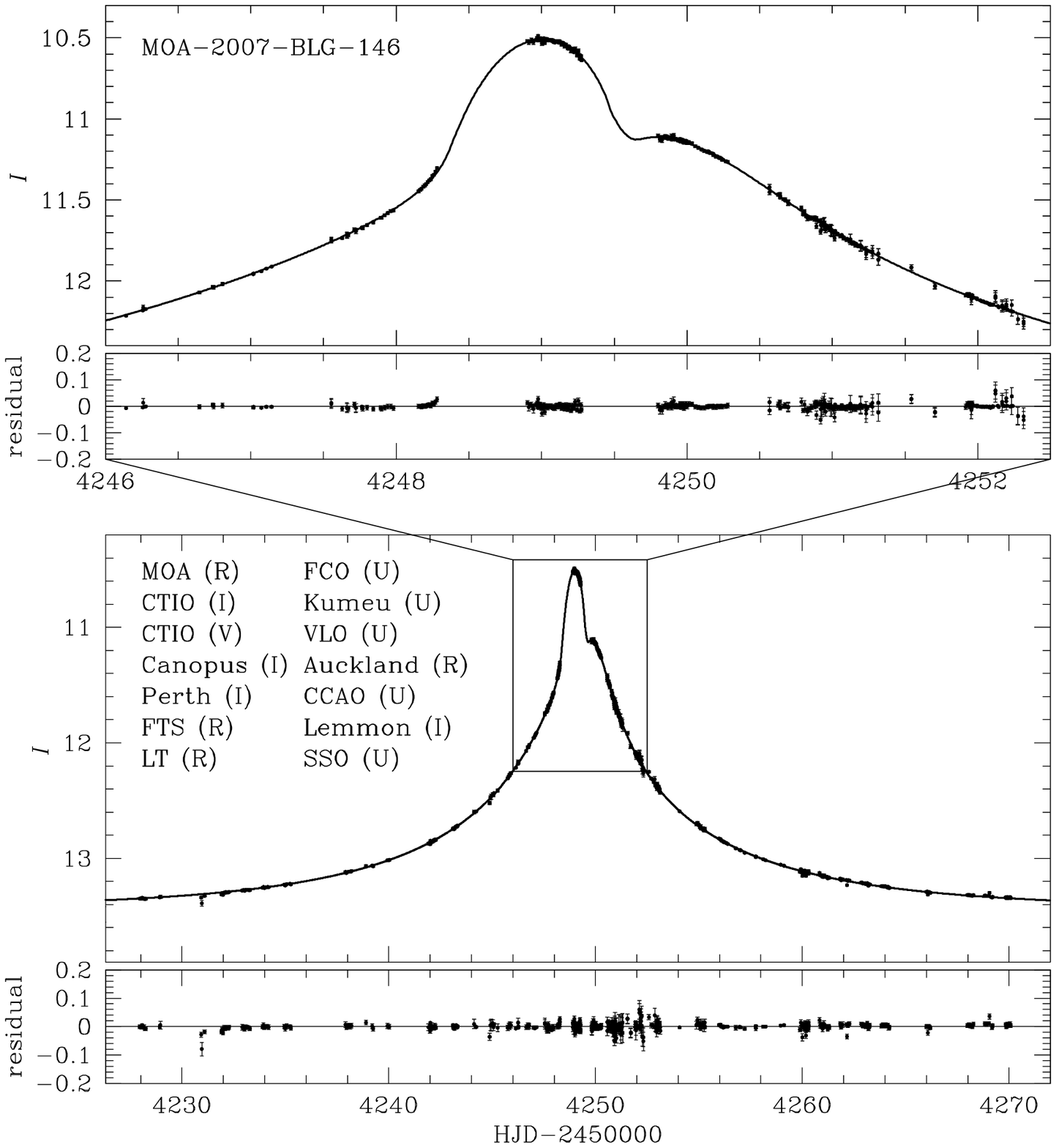}
\caption{\label{fig:one}
Light curve of the microlensing event MOA-2007-BLG-146.  The upper panel 
shows the enlargement of the region around the peak. The lensing parameters 
and the lens-system geometry corresponding to the best-fit model light curve 
are presented in Table \ref{table:five} and Fig.\ \ref{fig:ten}, respectively.
}\end{figure}

\begin{figure}[ht]
\epsscale{0.8}
\plotone{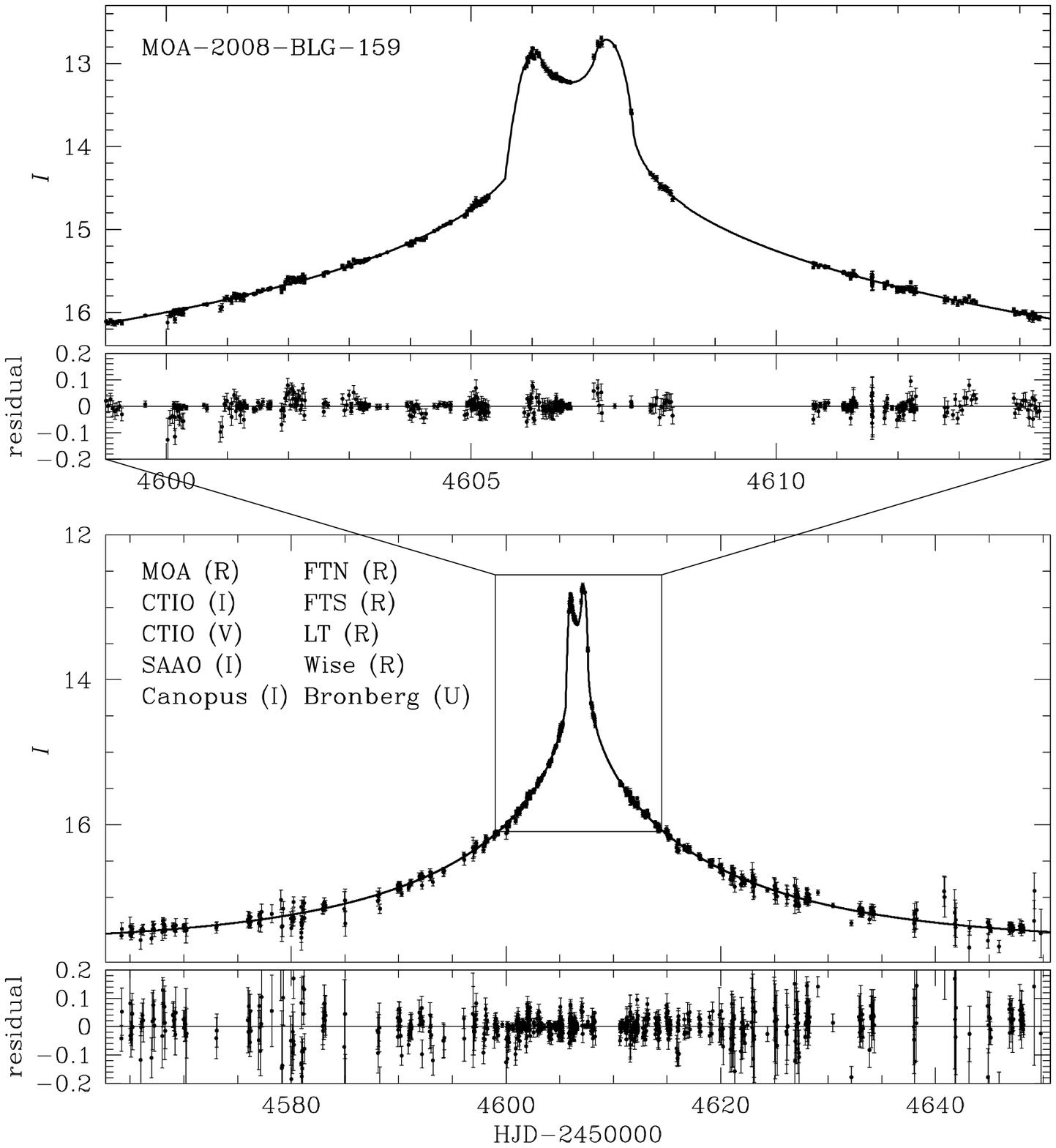}
\caption{\label{fig:two}
Light curve of the microlensing event MOA-2008-BLG-159.  
Notations same as in Fig.\ \ref{fig:one}.
}\end{figure}

\begin{figure}[ht]
\epsscale{0.8}
\plotone{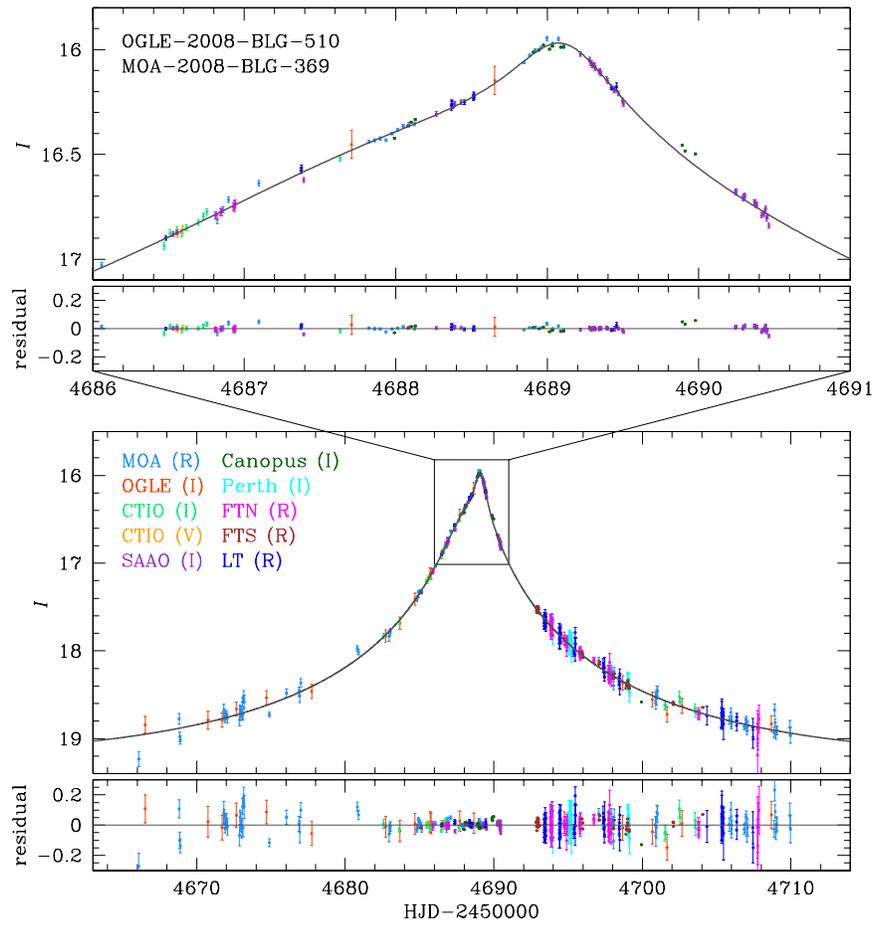}
\caption{\label{fig:three}
Light curve of the microlensing event OGLE-2008-BLG-510/MOA-2008-BLG-369.  
Notations same as in Fig.\ \ref{fig:one}.
}\end{figure}

\begin{figure}[ht]
\epsscale{0.8}
\plotone{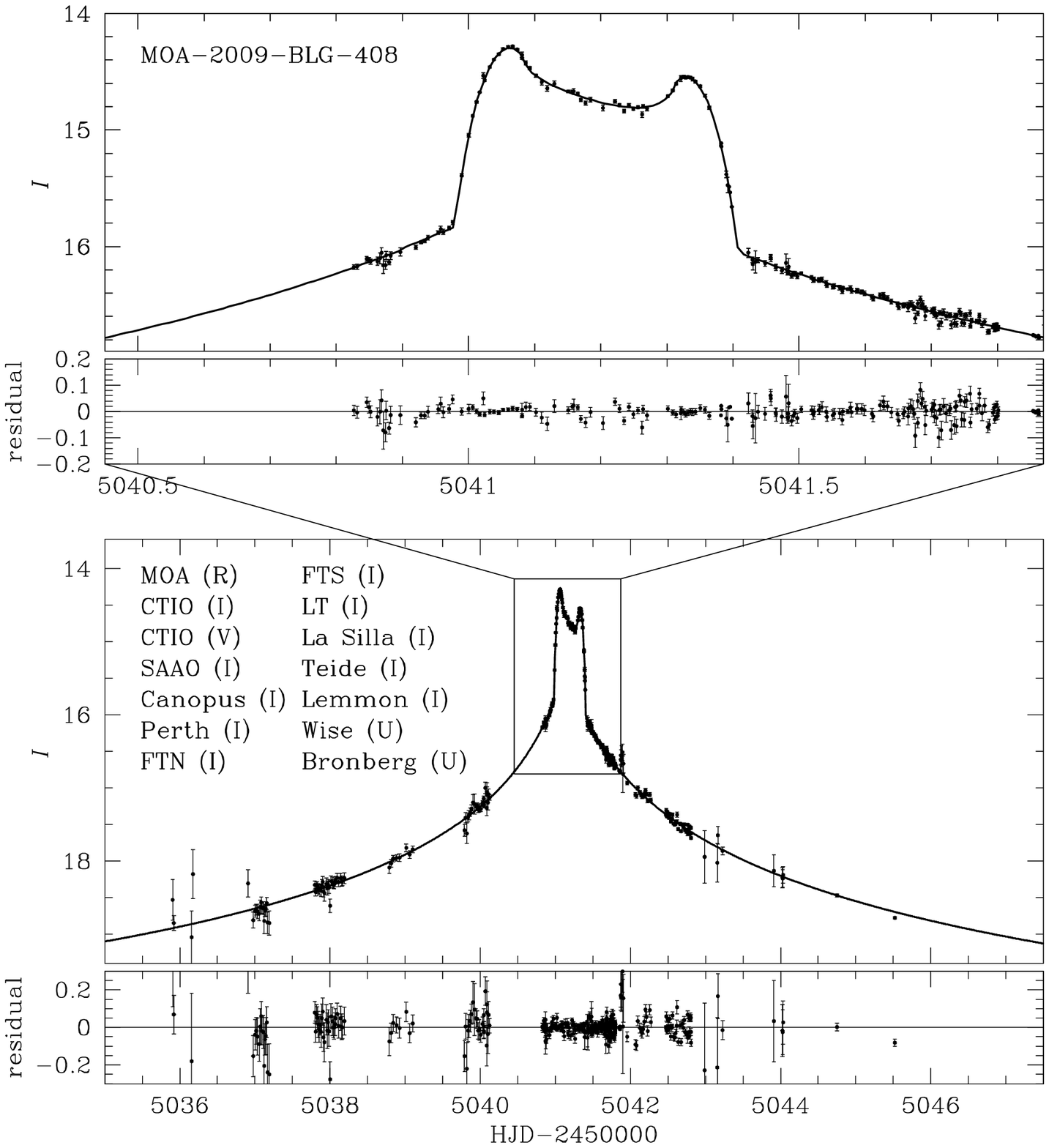}
\caption{\label{fig:four}
Light curve of the microlensing event MOA-2009-BLG-408.  
Notations same as in Fig.\ \ref{fig:one}.
}\end{figure}

\begin{figure}[ht]
\epsscale{0.8}
\plotone{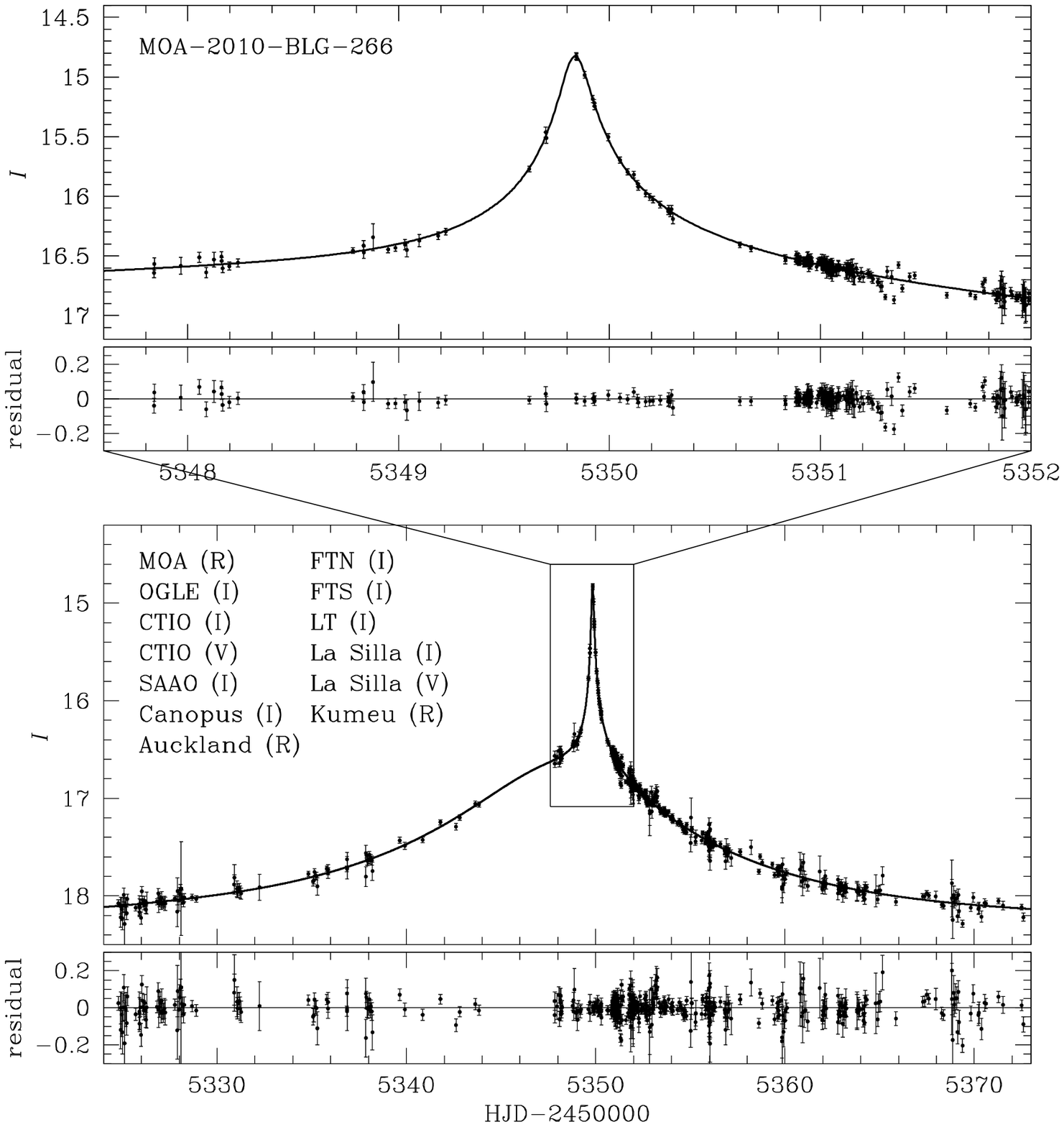}
\caption{\label{fig:five}
Light curve of the microlensing event MOA-2010-BLG-266.  
Notations same as in Fig.\ \ref{fig:one}.
}\end{figure}

\begin{figure}[ht]
\epsscale{0.8}
\plotone{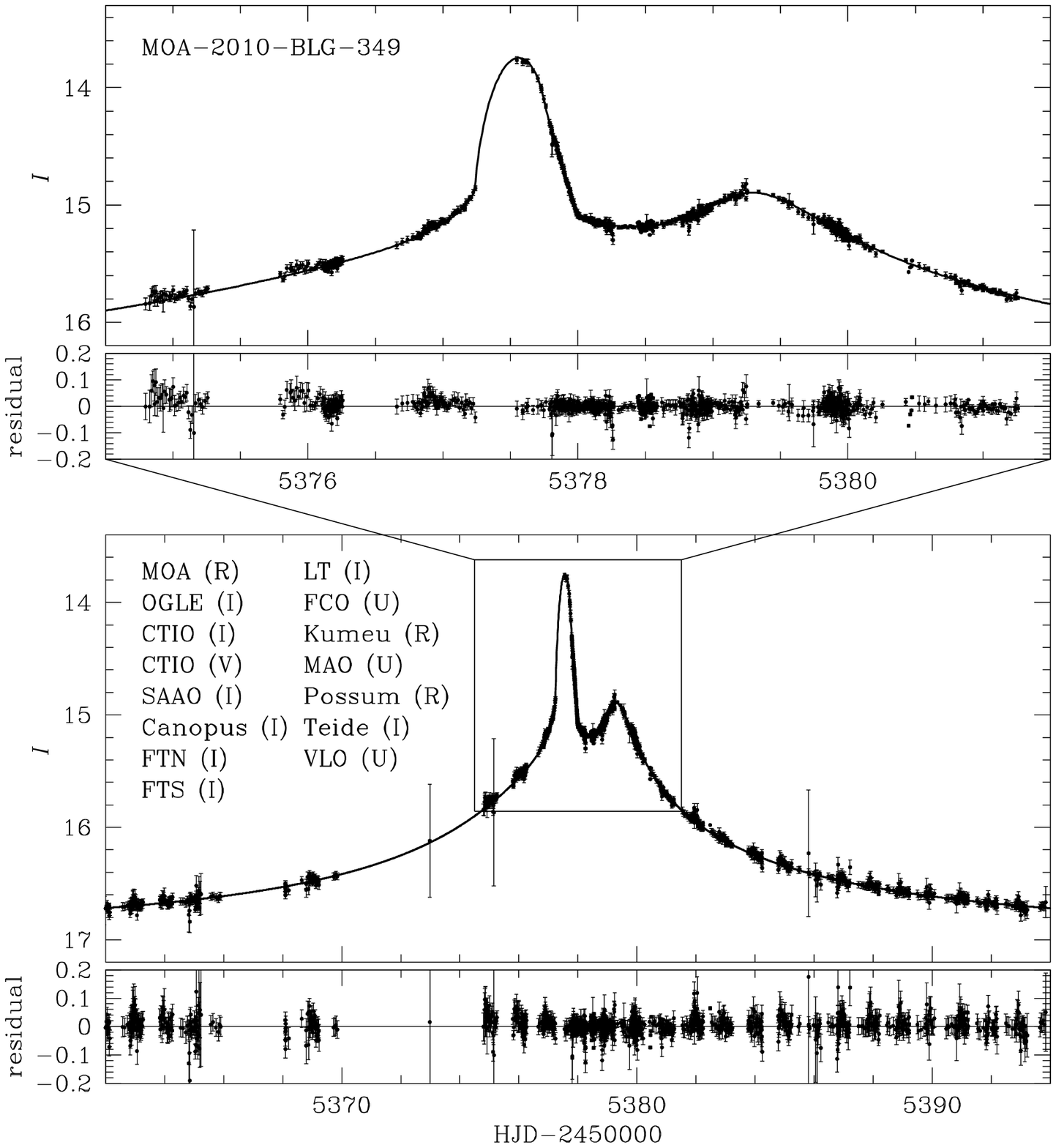}
\caption{\label{fig:six}
Light curve of the microlensing event MOA-2010-BLG-349.  
Notations same as in Fig.\ \ref{fig:one}.
}\end{figure}

\begin{figure}[ht]
\epsscale{0.8}
\plotone{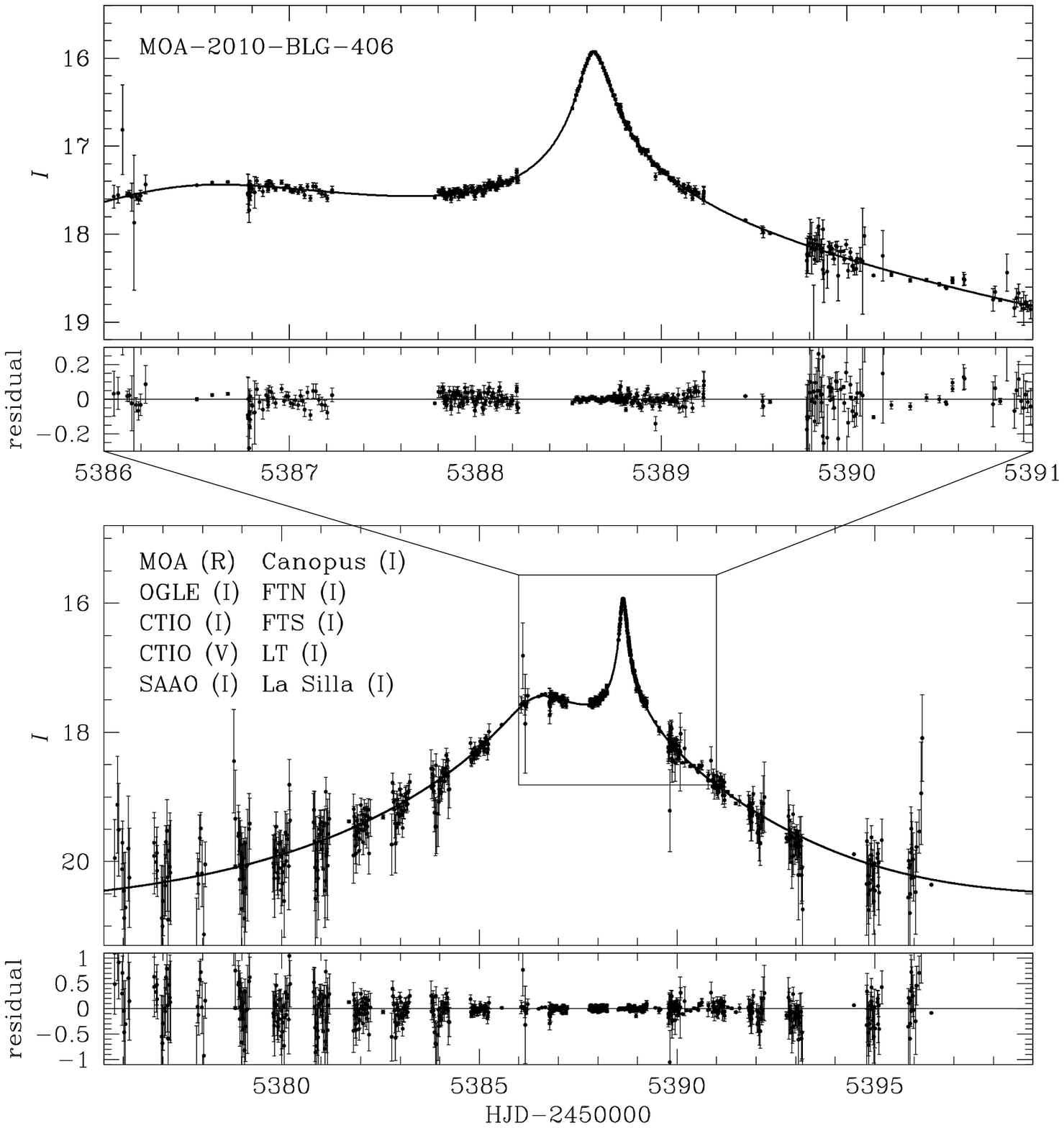}
\caption{\label{fig:seven}
Light curve of the microlensing event MOA-2010-BLG-406.  
Notations same as in Fig.\ \ref{fig:one}.
}\end{figure}

\begin{figure}[ht]
\epsscale{0.8}
\plotone{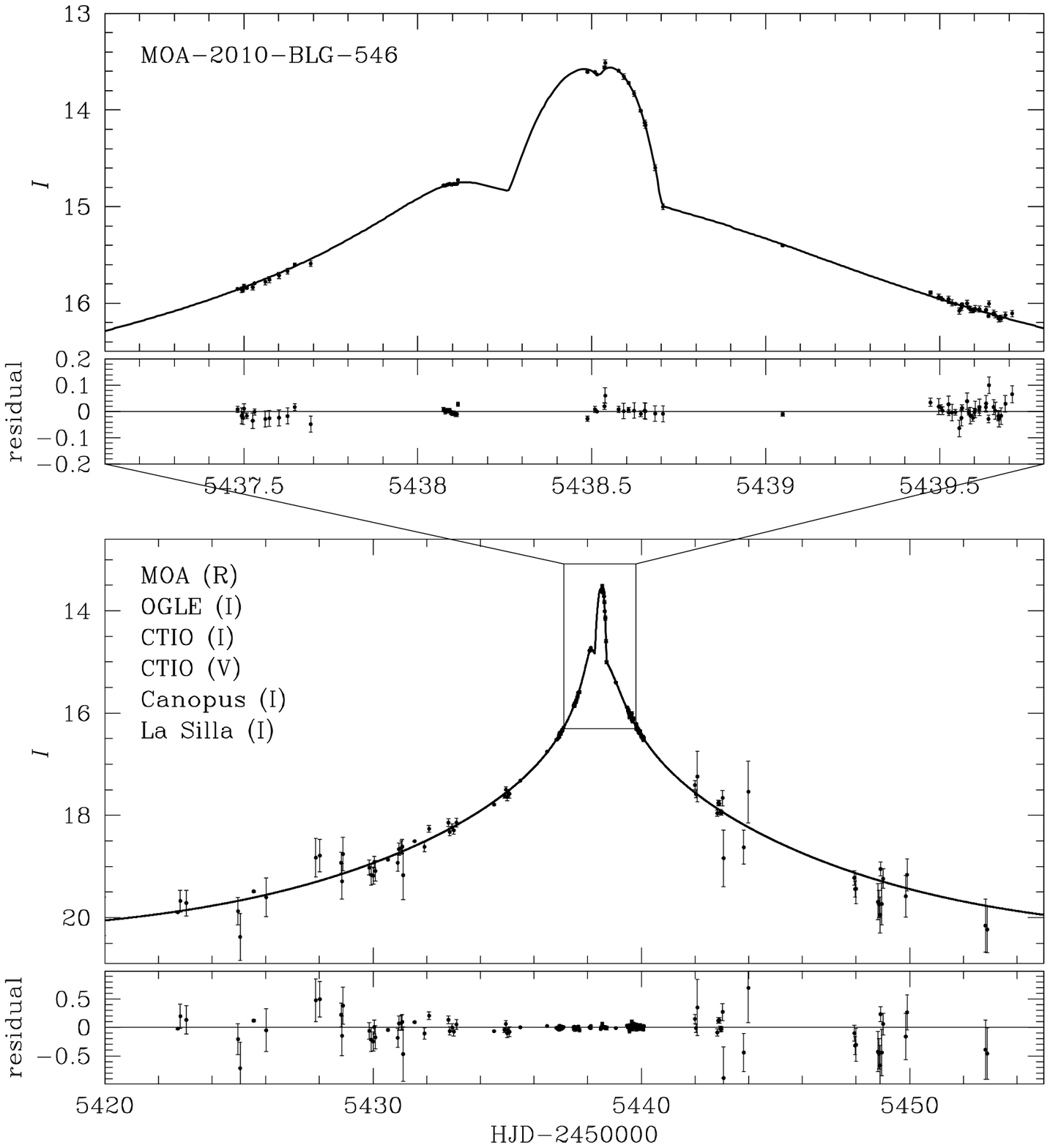}
\caption{\label{fig:eight}
Light curve of the microlensing event MOA-2010-BLG-546.  
Notations same as in Fig.\ \ref{fig:one}.
}\end{figure}

\begin{figure*}[ht]
\epsscale{1.0}
\plotone{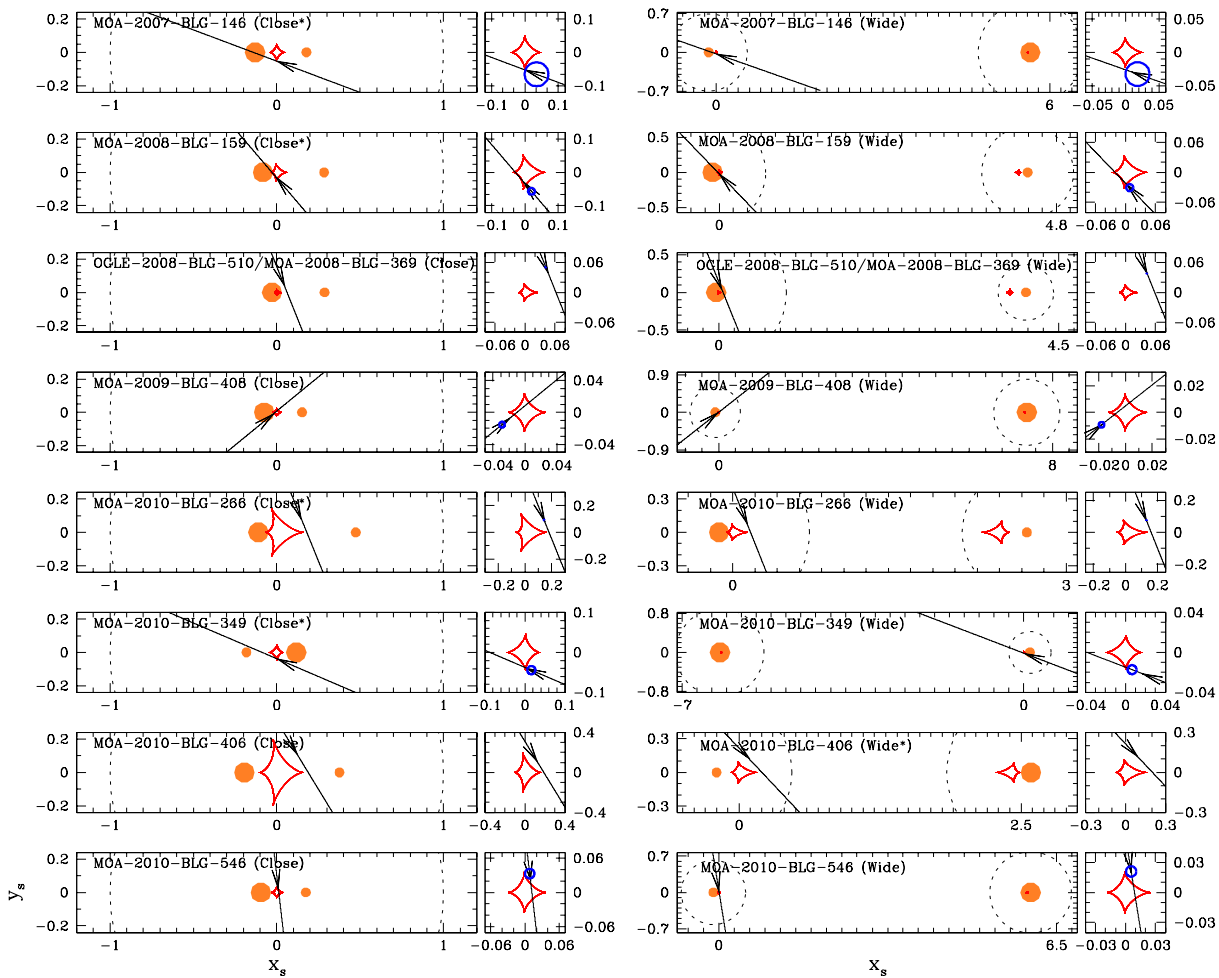}
\caption{\label{fig:nine}
Geometry of lens systems responsible for the light curves presented in 
Fig.\ 1 -- 8. For each event, we present two geometries corresponding to 
the close (left panels) and wide (right panels) binary solutions. 
The symbol `$\ast$' after the label `close' or `wide' indicates that 
the model is preferred over the other solution with 3$\sigma$ level.  In 
each panel, the big and small filled dots represent the lens components with 
heavier and lighter masses, respectively. The red closed figure represents 
the caustic and the straight line with an arrow is the source trajectory. 
The dashed circle represents the Einstein ring. For the close binary, there
is a single ring and its radius is the Einstein radius corresponding to the 
total mass of the binary. For the wide binary, on the other hand, there are 
two circles with their Einstein radii corresponding to the masses of the 
individual lens components. The small panel on the right side of each main 
panel shows the enlargement of the region around the caustic that caused 
perturbations. 
}\end{figure*}

\begin{figure}[ht]
\epsscale{0.8}
\plotone{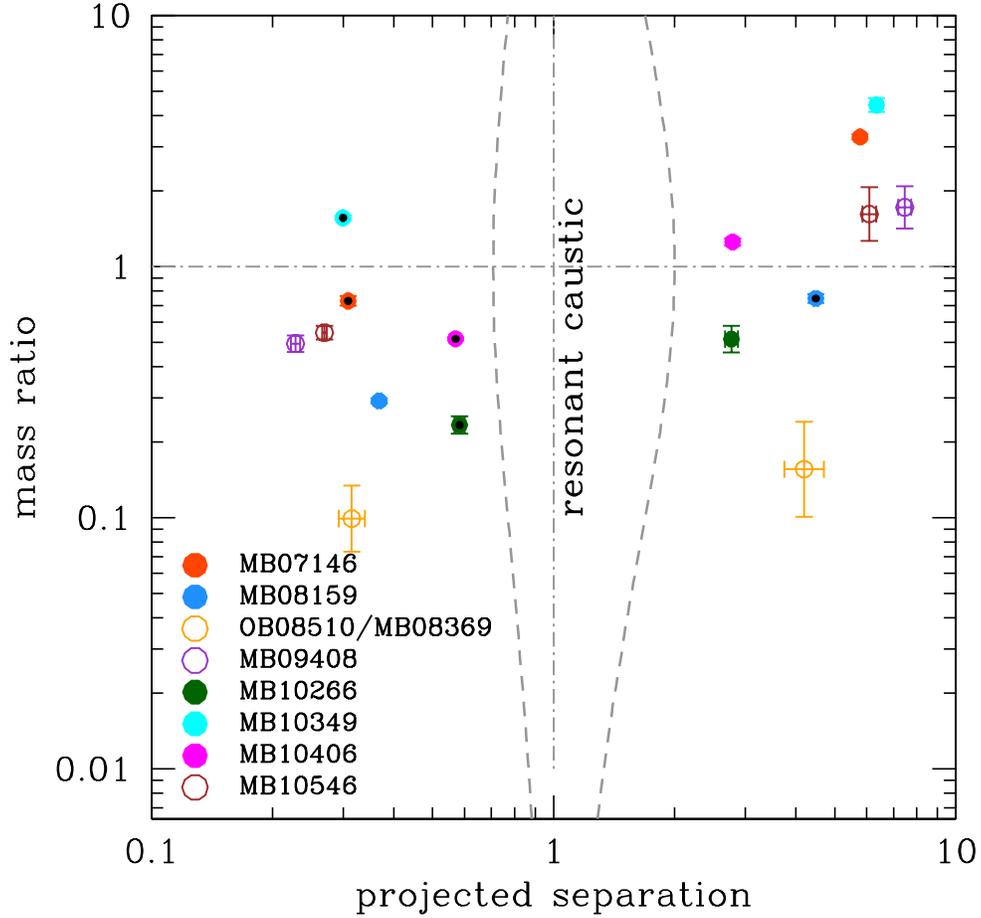}
\caption{\label{fig:ten}
Binary solutions in the parameter space of $(s,q)$.
The filled circles denote that the degeneracy is resolved with a $3\sigma$ 
confidence level and the empty circles symbolize the degeneracy is not resolved. 
Among a pair of solutions with resolved degeneracy, we mark a `$\bullet$'
sign inside a circle to indicate which solution is preferred.
The area encompassed by dashed lines represents the region within which the 
lens forms a single merged large caustic.
}\end{figure}

\headsep=30pt
\begin{sidewaystable}[ht]
\caption{Coordinates of Events\label{table:one}}
\begin{tabular}{lcccc}
\hline\hline 
\multicolumn{1}{c}{event} &
\multicolumn{1}{c}{RA} &
\multicolumn{1}{c}{DEC} &
\multicolumn{1}{c}{$l$} &
\multicolumn{1}{c}{$b$} \\
\hline
MOA-2007-BLG-146                   & 18$^{\rm h}$14$^{\rm m}$47$^{\rm s}$\hskip-2pt.72  & -27$^\circ$57'26"\hskip-2pt.9 & 04$^\circ$01'59"\hskip-2pt.23   & -05$^\circ$05'03"\hskip-2pt.08 \\
MOA-2008-BLG-159                   & 18$^{\rm h}$07$^{\rm m}$29$^{\rm s}$\hskip-2pt.18  & -30$^\circ$09'49"\hskip-2pt.1 & 01$^\circ$19'22"\hskip-2pt.26   & -04$^\circ$43'46"\hskip-2pt.96 \\
OGLE-2008-BLG-510/MOA-2008-BLG-369 & 18$^{\rm h}$09$^{\rm m}$37$^{\rm s}$\hskip-2pt.65  & -26$^\circ$02'26"\hskip-2pt.7 & 05$^\circ$10'23"\hskip-2pt.63   & -03$^\circ$09'32"\hskip-2pt.98 \\
MOA-2009-BLG-408                   & 17$^{\rm h}$57$^{\rm m}$08$^{\rm s}$\hskip-2pt.01  & -30$^\circ$44'18"\hskip-2pt.4 & 359$^\circ$43'27"\hskip-2pt.80  & -03$^\circ$04'00"\hskip-2pt.48 \\
MOA-2010-BLG-266                   & 17$^{\rm h}$54$^{\rm m}$50$^{\rm s}$\hskip-2pt.84  & -34$^\circ$15'40"\hskip-2pt.4 & 356$^\circ$25'32"\hskip-2pt.88  & -04$^\circ$24'41"\hskip-2pt.22 \\
MOA-2010-BLG-349                   & 17$^{\rm h}$53$^{\rm m}$27$^{\rm s}$\hskip-2pt.65  & -28$^\circ$24'43"\hskip-2pt.3 & 01$^\circ$20'01"\hskip-2pt.50   & -01$^\circ$12'20"\hskip-2pt.74 \\
MOA-2010-BLG-406                   & 17$^{\rm h}$55$^{\rm m}$27$^{\rm s}$\hskip-2pt.52  & -31$^\circ$38'55"\hskip-2pt.2 & 358$^\circ$45'20"\hskip-2pt.56  & -03$^\circ$12'44"\hskip-2pt.82 \\
MOA-2010-BLG-546                   & 17$^{\rm h}$59$^{\rm m}$57$^{\rm s}$\hskip-2pt.69  & -31$^\circ$35'32"\hskip-2pt.5 & 359$^\circ$16'57"\hskip-2pt.50  & -04$^\circ$00'57"\hskip-2pt.13 \\
\hline
\end{tabular}
\end{sidewaystable}

\headsep=30pt
\begin{table}[ht]
\caption{Observatories\label{table:two}}
\begin{tabular}{lllllll}
\hline\hline 
\multicolumn{1}{c}{event} &
\multicolumn{1}{c}{MOA} &
\multicolumn{1}{c}{OGLE} &
\multicolumn{1}{c}{$\mu$FUN} &
\multicolumn{1}{c}{PLANET} &
\multicolumn{1}{c}{RoboNet} &
\multicolumn{1}{c}{MiNDSTEp} \\
\hline
MOA-2007-BLG-146   & Mt. John  &              & CTIO       & Canopus & FTS  &           \\
                   &           &              & Auckland   & Perth   & LT   &           \\
                   &           &              & CCAO       &         &      &           \\
                   &           &              & FCO        &         &      &           \\
                   &           &              & Kumeu      &         &      &           \\
                   &           &              & Lemmon     &         &      &           \\
                   &           &              & SSO        &         &      &           \\
                   &           &              & VLO        &         &      &           \\
\hline
MOA-2008-BLG-159   & Mt. John  &              & CTIO       & SAAO    & FTN  &           \\
                   &           &              & Wise       & Canopus & FTS  &           \\
                   &           &              & Bronberg   &         & LT   &           \\
\hline
OGLE-2008-BLG-510/ & Mt. John  & LCO          & CTIO       & SAAO    & FTN  &           \\ 
MOA-2008-BLG-369   &           &              &            & Canopus & FTS  &           \\
                   &           &              &            & Perth   & LT   &           \\   
\hline
MOA-2009-BLG-408   & Mt. John  &              & CTIO       & SAAO    & FTN  & La Silla  \\
                   &           &              & Wise       & Canopus & FTS  &           \\   
                   &           &              & Bronberg   & Perth   & LT   &           \\   
                   &           &              & Lemmon     &         &      &           \\   
                   &           &              & Teide      &         &      &           \\   
\hline
MOA-2010-BLG-266   & Mt. John  & LCO          & CTIO       & SAAO    & FTN  & La Silla  \\
                   &           &              & Auckland   & Canopus & FTS  &           \\   
                   &           &              & Kumeu      &         & LT   &           \\   
\hline
MOA-2010-BLG-349   & Mt. John  & LCO          & CTIO       & SAAO    & FTN  &           \\
                   &           &              & FCO        & Canopus & FTS  &           \\   
                   &           &              & Kumeu      &         & LT   &           \\   
                   &           &              & MAO        &         &      &           \\   
                   &           &              & Possum     &         &      &           \\   
                   &           &              & Teide      &         &      &           \\   
                   &           &              & VLO        &         &      &           \\   
\hline
MOA-2010-BLG-406   & Mt. John  & LCO          & CTIO       & SAAO    & FTN  & La Silla  \\
                   &           &              &            & Canopus & FTS  &           \\   
                   &           &              &            &         & LT   &           \\   
\hline
MOA-2010-BLG-546   & Mt. John  & LCO          & CTIO       & Canopus &      & La Silla  \\
\hline
\end{tabular}
\\LCO: Las Campanas Observatory; 
CTIO: Cerro Tololo Inter-American Observatory;
CCAO: Campo Catino Austral Observatory;
FCO: Farm Cove Observatory;
SSO: Southern Stars Observatory;
VLO: Vintage Lane Observatory;
MAO: Molehill Astronomical Observatory;
SAAO: South Africa Astronomy Astronomical Observatory;
FTN: Faulkes North;
FTS: Faulkes South;
LT: Liverpool Telescope.
\end{table}

\headsep=30pt
\begin{table}[ht]
\caption{Telescopes\label{table:three}}
\begin{tabular}{ll}
\hline\hline 
\multicolumn{1}{c}{telescope} &
\multicolumn{1}{c}{location} \\
\hline
MOA 2.0 m Mt. John       & New Zealand           \\
OGLE 1.3 m Warsaw        & Las Campanas, Chile   \\
$\mu$FUN 1.3 m SMART     & CTIO Chile            \\
$\mu$FUN 0.4 m Auckland  & New Zealand           \\
$\mu$FUN 0.4 m CCAO      & Chile                 \\
$\mu$FUN 0.4 m FCO       & New Zealand           \\
$\mu$FUN 0.4 m Kumeu     & New Zealand           \\
$\mu$FUN 1.0 m Lemmon    & Arizona               \\
$\mu$FUN 0.4 m VLO       & New Zealand           \\
$\mu$FUN 0.5 m Wise      & Israel                \\
$\mu$FUN 0.4 m Bronberg  & South Africa          \\
$\mu$FUN 0.8 m Teide     & Canary Islands, Spain \\
$\mu$FUN 0.3 m MAO       & New Zealand           \\
$\mu$FUN 0.4 m Possum    & New Zealand           \\
$\mu$FUN 0.3 m SSO       & Tahiti                \\
PLANET 1.0 m SAAO        & South Africa          \\
PLANET 1.0 m Canopus     & Australia             \\
PLANET 0.6 m Perth       & Australia             \\
RoboNet 2.0 m FTN        & Hawaii                \\
RoboNet 2.0 m FTS        & Australia             \\
RoboNet 2.0 m LT         & La Palma, Spain       \\
MiNDSTEp 1.54 m Danish   & La Silla, Chile       \\
\hline
\end{tabular}
\end{table}

\headsep=30pt
\begin{table}[ht]
\caption{Limb-darkening Coefficients\label{table:four}}
\begin{tabular}{lcccc}
\hline\hline 
\multicolumn{1}{c}{event} &
\multicolumn{1}{c}{$\Gamma_V$} &
\multicolumn{1}{c}{$\Gamma_R$} &
\multicolumn{1}{c}{$\Gamma_I$} \\
\hline
MOA-2007-BLG-146                    & 0.74 &  0.64 &  0.53 \\
MOA-2008-BLG-159                    & 0.57 &  0.48 &  0.40 \\
OGLE-2008-BLG-510/MOA-2008-BLG-369  & --   &  --   &  --   \\
MOA-2009-BLG-408                    & 0.65 &  0.56 &  0.47 \\
MOA-2010-BLG-266                    & --   &  --   &  --   \\
MOA-2010-BLG-349                    & 0.65 &  0.58 &  0.48 \\
MOA-2010-BLG-406                    & --   &  --   &  --   \\
MOA-2010-BLG-546                    & 0.68 &  0.59 &  0.49 \\
\hline
\end{tabular}
\end{table}

\headsep=30pt
\begin{sidewaystable}[ht]
\caption{Best-fit Model Parameters\label{table:five}}
\begin{tabular}{lllllllllllll}
\hline\hline 
\multicolumn{1}{c}{event} &
\multicolumn{1}{c}{model} &
\multicolumn{1}{c}{$\chi^2/{\rm dof}$} &
\multicolumn{1}{c}{$t_0$} &
\multicolumn{1}{c}{$u_0$} &
\multicolumn{1}{c}{$t_{\rm E}$} &
\multicolumn{1}{c}{$s$} &
\multicolumn{1}{c}{$q$} &
\multicolumn{1}{c}{$\alpha$} &
\multicolumn{1}{c}{$\rho_\star$} &
\multicolumn{1}{c}{$\theta_{\star}$} &
\multicolumn{1}{c}{$\theta_{\rm E}$} &
\multicolumn{1}{c}{$\mu$}\\
\multicolumn{1}{c}{} &
\multicolumn{1}{c}{} &
\multicolumn{1}{c}{} &
\multicolumn{1}{c}{(HJD')} &
\multicolumn{1}{c}{} &
\multicolumn{1}{c}{(days)} &
\multicolumn{1}{c}{} &
\multicolumn{1}{c}{} &
\multicolumn{1}{c}{} &
\multicolumn{1}{c}{} &
\multicolumn{1}{c}{($\mu$as)} &
\multicolumn{1}{c}{(mas)} &
\multicolumn{1}{c}{(mas/yr)} \\
\hline
MOA-2007-BLG-146  & close & 1550.7 & 4249.16    & 0.049      & 15.506     & 0.308      & 0.729      & 3.501      & 0.036      & 15.512     & 0.435      & 10.237     \\
                  &       & /1560  & $\pm$0.004 & $\pm$0.001 & $\pm$0.077 & $\pm$0.002 & $\pm$0.032 & $\pm$0.004 & $\pm$0.001 & $\pm$1.343 & $\pm$0.040 & $\pm$0.932 \\
                  & wide  & 1855.6 & 4249.13    & 0.053      & 14.081     & 5.785      & 3.279      & 3.480      & 0.038      & 16.013     & 0.433      & 10.960     \\
                  &       & /1560  & $\pm$0.003 & $\pm$0.001 & $\pm$0.070 & $\pm$0.025 & $\pm$0.077 & $\pm$0.001 & $\pm$0.001 & $\pm$1.387 & $\pm$0.040 & $\pm$0.998 \\
\hline
MOA-2008-BLG-159  & close & 2407.3 & 4606.74    & 0.022      & 29.180     & 0.368      & 0.292      & 4.006      & 0.010      & 1.588      & 0.156      & 1.950      \\
                  &       & /2418  & $\pm$0.004 & $\pm$0.001 & $\pm$0.343 & $\pm$0.004 & $\pm$0.007 & $\pm$0.004 & $\pm$0.001 & $\pm$0.137 & $\pm$0.020 & $\pm$0.255 \\
                  & wide  & 2472.1 & 4606.66    & 0.020      & 32.221     & 4.486      & 0.747      & 3.947      & 0.009      & 1.545      & 0.169      & 1.911      \\
                  &       & /2418  & $\pm$0.005 & $\pm$0.001 & $\pm$0.379 & $\pm$0.056 & $\pm$0.029 & $\pm$0.004 & $\pm$0.001 & $\pm$0.134 & $\pm$0.022 & $\pm$0.250 \\
\hline
OGLE-2008-BLG-510 & close & 1879.2 & 4688.67    & 0.057      & 21.531     & 0.315      & 0.099      & 1.191      & --         & --         & --         & --        \\
/MOA-2008-BLG-369 &       & /1918  & $\pm$0.007 & $\pm$0.002 & $\pm$0.641 & $\pm$0.023 & $\pm$0.030 & $\pm$0.007 & --         & --         & --         & --        \\
                  & wide  & 1878.1 & 4688.65    & 0.058      & 21.972     & 4.100      & 0.156      & 1.187      & --         & --         & --         & --        \\
                  &       & /1918  & $\pm$0.006 & $\pm$0.002 & $\pm$0.654 & $\pm$0.471 & $\pm$0.068 & $\pm$0.008 & --         & --         & --         & --        \\
\hline
MOA-2009-BLG-408  & close & 1740.8 & 5041.20    & 0.006      & 13.769     & 0.228      & 0.493      & 5.597      & 0.004      & 0.955      & 0.263      & 6.975      \\
                  &       & /1729  & $\pm$0.002 & $\pm$0.001 & $\pm$0.543 & $\pm$0.006 & $\pm$0.037 & $\pm$0.009 & $\pm$0.001 & $\pm$0.083 & $\pm$0.076 & $\pm$2.013 \\
                  & wide  & 1740.0 & 5041.20    & 0.007      & 13.886     & 7.472      & 1.720      & 5.616      & 0.003      & 0.946      & 0.266      & 6.994      \\
                  &       & /1729  & $\pm$0.002 & $\pm$0.001 & $\pm$0.548 & $\pm$0.280 & $\pm$0.332 & $\pm$0.008 & $\pm$0.001 & $\pm$0.082 & $\pm$0.077 & $\pm$2.018 \\
\hline
MOA-2010-BLG-266  & close & 4817.0 & 5348.85    & 0.167      & 14.632     & 0.583      & 0.234      & 1.186      & --         & --      & --         & --        \\
                  &       & /4818  & $\pm$0.054 & $\pm$0.008 & $\pm$0.324 & $\pm$0.017 & $\pm$0.019 & $\pm$0.012 & --         & --      & --         & --        \\
                  & wide  & 4837.4 & 5348.70    & 0.183      & 15.702     & 2.768      & 0.514      & 1.191      & --         & --      & --         & --        \\
                  &       & /4818  & $\pm$0.032 & $\pm$0.009 & $\pm$0.348 & $\pm$0.099 & $\pm$0.063 & $\pm$0.013 & --         & --      & --         & --        \\
\hline
\end{tabular}
\end{sidewaystable}

\headsep=30pt
\begin{sidewaystable}[ht]
\caption{Table 5 continued}
\begin{tabular}{lllllllllllll}
\hline\hline 

\hline
MOA-2010-BLG-349  & close & 7883.4 & 5377.92    & 0.034      & 24.695     & 0.299      & 1.562      & 3.546      & 0.010      & 4.713      & 0.458      & 6.775      \\
                  &       & /7946  & $\pm$0.003 & $\pm$0.001 & $\pm$0.193 & $\pm$0.001 & $\pm$0.034 & $\pm$0.002 & $\pm$0.001 & $\pm$0.408 & $\pm$0.060 & $\pm$0.882 \\
                  & wide  & 7909.6 & 5377.85    & 0.033      & 24.530     & 6.351      & 4.391      & 0.365      & 0.009      & 4.617      & 0.443      & 6.593      \\
                  &       & /7946  & $\pm$0.003 & $\pm$0.001 & $\pm$0.192 & $\pm$0.029 & $\pm$0.278 & $\pm$0.002 & $\pm$0.001 & $\pm$0.400 & $\pm$0.058 & $\pm$0.858 \\
\hline
MOA-2010-BLG-406  & close & 2108.9 & 5388.13    & 0.161      & 5.359      & 0.570      & 0.515      & 1.024      & --         & --      & --         & --        \\
                  &       & /2030  & $\pm$0.005 & $\pm$0.001 & $\pm$0.057 & $\pm$0.002 & $\pm$0.007 & $\pm$0.004 & --         & --      & --         & --        \\
                  & wide  & 2020.4 & 5387.53    & 0.221      & 5.362      & 2.787      & 1.252      & 0.815      & --         & --      & --         & --        \\
                  &       & /2030  & $\pm$0.007 & $\pm$0.004 & $\pm$0.083 & $\pm$0.018 & $\pm$0.039 & $\pm$0.003 & --         & --      & --         & --        \\
\hline
MOA-2010-BLG-546  & close & 462.1  & 5438.49    & 0.012      & 8.814      & 0.269      & 0.546      & 1.455      & 0.008      & 1.822      & 0.219      & 9.082      \\
                  &       & /458   & $\pm$0.003 & $\pm$0.001 & $\pm$0.164 & $\pm$0.004 & $\pm$0.035 & $\pm$0.008 & $\pm$0.001 & $\pm$0.158 & $\pm$0.033 & $\pm$1.347 \\
                  & wide  & 458.4  & 5438.50    & 0.015      & 9.305      & 6.102      & 1.618      & 1.411      & 0.008      & 1.777      & 0.214      & 8.396      \\
                  &       & /458   & $\pm$0.002 & $\pm$0.001 & $\pm$0.173 & $\pm$0.237 & $\pm$0.398 & $\pm$0.009 & $\pm$0.001 & $\pm$0.154 & $\pm$0.032 & $\pm$1.245 \\
\hline
\end{tabular}
\\$\rm  HJD'=HJD-2450000$. For the wide binary solutions, the lensing parameters 
$u_0$ and $\rho_\star$ are normalized by the radius of the Einstein radius 
corresponding to the mass of the binary lens component that the source trajectory
approaches close to. The Einstein time scale, $t_{\rm E}$, and the Einstein radius, 
$\theta_{\rm E}$, are similarly normalized.  We also note that $q<1$ and $q>1$ 
represent the cases where the source trajectory approaches the heavier and lighter 
lens components, respectively. The Einstein radius is determined by $\theta_{\rm E}
=\theta_\star/\rho_\star$ where the angular radius of the source star $\theta_\star$ 
is measured based on the source brightness and color.  For events where the 
perturbations do not result from caustic crossings, the values of $\rho_\star$ and 
$\theta_{\rm E}$ cannot be measured and thus are not presented.
\end{sidewaystable}


\begin{thebibliography}{99}

\bibitem[Albrow et al.(2002)]{albrow02}
Albrow, M.\ D., et al.\ 2002, \apj, 572, 1031

\bibitem[Alcock et al.(1995)]{alcock95}
Alcock, C., et al. 1995, \apj, 454, L125

\bibitem[Alcock et al.(2000)]{alcock00}
Alcock, C., et al.\ 2000, \apj, 541, 270

\bibitem[Beaulieu et al.(2006)]{beaulieu06}
Beaulieu, J.-P., et al.\ 2006, Nature, 439, 437

\bibitem[Bond et al.(2001)]{bond01}
Bond, I. A., et al.\ 2001, \mnras, 327, 868

\bibitem[Bozza et al.(2011)]{bozza11}
Bozza, V., et al.\ 2011, \mnras, submitted

\bibitem[Claret(2000)]{claret00}
Claret, A.\ 2000, \aap, 363, 1081

\bibitem[Dominik(1999)]{dominik99}
Dominik, M.\ 1999, \aap, 349, 108

\bibitem[Dominik et al.(2007)]{dominik07}
Dominik, M., et al. 2007, MNRAS, 380, 792

\bibitem[Dominik et al.(2010)]{dominik10}
Dominik, M., et al.\ 2010, Astron.\ Machr., 331, 671

\bibitem[Dong et al.(2006)]{dong06}
Dong, S., et al.\ 2009 \apj, 642, 842

\bibitem[Gould(1992)]{gould92}
Gould, A.\ 1992, \apj, 392, 442



\bibitem[Gould(2001)]{gould01}
Gould, A.\ 2001, PASP, 113, 903

\bibitem[Gould(2008)]{gould08}
Gould, A.\ 2008, \apj, 681, 1593

\bibitem[Gould et al.(2006)]{gould06}
Gould, A., et al.\ 2006, \apj, 644, L37

\bibitem[Griest \& Safizadeh(1998)]{griest98}
Griest, K., \& Safizadeh, N.\ 1998, \apj, 500, 37

\bibitem[Han et al. (1999)]{han99}
Han, C., Chun, M -S., \& Chang, K. 1999, \apj, 536, 406

\bibitem[Han \& Gould (2000)]{han00}
Han, C., \& Gould, A. 2000, \apj, 538, 653

\bibitem[Han (2009)]{han09a}
Han, C.\ 2009, \apj, 691, L9

\bibitem[Han (2009)]{han09b}
Han, C.\ 2009, \apj, 707, 1264

\bibitem[Han \& Gaudi (2008)]{han08}
Han, C., \& Gaudi B.\ S.\ 2008, \apj, 689, 53

\bibitem[Han \& Kim(2009)]{hankim09}
Han, C., \& Kim, D.\ 2009, \apj, 693, 1835

\bibitem[Jaroszy\'nski, et al.(2004)]{jaroszynski04}
Jaroszy\'nski, M., et al.\ 2004, Acta Astron., 54, 103

\bibitem[Jaroszy\'nski, et al.(2006)]{jaroszynski06}
Jaroszy\'nski, M., et al.\ 2006, Acta Astron., 56, 307

\bibitem[Jaroszy\'nski, et al.(2010)]{jaroszynski10}
Jaroszy\'nski, M., et al.\ 2010, Acta Astron., 60, 197

\bibitem[Kayser et al.(1986)]{kayser86}
Kayser, R., Refsdal, S., \& Stabell, R.\ 1986, \aap, 166, 36

\bibitem[Paczy\'nski(1986)]{paczynski86}
Paczy\'nski, B.\ 1986, \apj, 304, 1

\bibitem[Pejcha \& Heyrovsk\'y(2009)]{pejcha09}
Pejcha, O., \& Heyrovsk\'y, D.\ 2009, \apj, 690, 1772

\bibitem[Ryu et al.(2010)]{ryu10}
Ryu, Y.-H., et al. 2010, \apj, 723, 81

\bibitem[Schneider \& Weiss(1986)]{schneider86}
Schneider, P., \& Weiss, A.\ 1986, \aap, 164, 237

\bibitem[Shin et al.(2011)]{shin11}
Shin, I.-G., et al. 2011, \apj, 735, 855

\bibitem[Skowron et al.(2007)]{skowron07}
Skowron, J., et al.\ 2007, Acta Astron., 57, 281

\bibitem[Skowron et al.(2011)]{skowron11}
Skowron, J., et al.\ 2011, \apj, 738, 87

\bibitem[Sumi et al.(2003)]{sumi03}
Sumi, T., et al.\ 2003, \apj, 591, 204

\bibitem[Tsapras et al.(2009)]{tsapras09}
Tsapras, Y., et al.\ 2009, Astron. Nachr., 330, 4

\bibitem[Udalski et al.(1994)]{udalski94}
Udalski, A., Szyma{\'n}ski, M., Mao, S., Di Stefano, R., Ka{\l}u{\.z}ny, J., Kubiak, M.,
Mateo, M., \& Krzemi{\'n}ski, W.\ 1994, \apj, 436, L103 

\bibitem[Udalski(2003)]{udalski03}
Udalski, A.\ 2003, Acta Astron., 53, 291

\bibitem[Yoo et al.(2004)]{yoo04}
Yoo, J., et al.\ 2004, \apj, 603, 139

\bibitem[Wambsganss(1997)]{wambsganss97}
Wambsganss, J.\ 1997, \mnras, 284, 172

\end{thebibliography}
\end{document}